# Magnetic reversal processes and critical thickness in FePt/α-Fe/FePt trilayers

N. L. Guo [1], G. P. Zhao [1-3*], H. W. Zhang [2], Y. Deng [1]

[1] College of Physics and Electronic Engineering, Sichuan Normal University, Chengdu 610068, China

[2] State Key Laboratory of Electronic Thin Films and Integrated Devices, University of Electronic Science and Technology of China, Chengdu 610054, China

[3] National Laboratory of Solid State Microstructures, Nanjing University, Nanjing 210093, China

**Abstract**

Magnetic reversal processes of a FePt/α-Fe/FePt trilayer system with in-plane easy axes have been investigated within a micromagnetic approach. It is found that the magnetic reversal process consists of three steps: nucleation of a prototype of domain wall in the soft phase, the evolution as well as the motion of the domain wall from the soft to the hard phase and finally, the magnetic reversal of the hard phase. For small soft layer thickness $L^s$, the three steps are reduced to one single step, where the magnetizations in the two phases reverse simultaneously and the hysteresis loops are square with nucleation as the coercivity mechanism. As $L^s$ increases, both nucleation and pinning fields decrease. In the meantime, the single-step reversal expands to a standard three-step one and the coercivity mechanism changes from nucleation to pinning. The critical thickness where the coercivity mechanism alters, could be derived analytically, which is found to be inversely proportional to the square root of the crystalline anisotropy of the hard phase. Further increase of $L^s$ leads to the change of the coercivity mechanism from

---

\* † E-mail: zhaogp@uestc.edu.cn

pinning to nucleation

## 1. Introduction

Magnetic multilayers, with importance in both theory and application, have been an intensive topic[1-17] in recent years. The hard/soft multilayers, which combine high remanence of the soft phase and high coercivity of the hard phase, are thought of as the best permanent magnets. Many works have been done on these materials to achieve the giant energy products predicted by Skomski and Coey[2] in 1993. Among them, the micromagnetic method is one of the most important theoretical methods.

Many theoretical works in this field focus on the macroscopic behaviors of the hysteresis loops. The calculated coercivity and energy products are still much larger than those realized in experiments. The microscopic hysteresis loops, which could give clear magnetic reversal process and coercivity mechanism, have been seldom investigated.

In this paper, the magnetic reversal processes in parallel-oriented magnetic trilayers of FePt/α-Fe have been investigated systematically within a self-contained micromagnetic approach, with both micromagnetic and macroscopic hysteresis loops obtained numerically. In particular, the critical thickness, at which the coercivity mechanism changes, has been derived analytically along with the nucleation field.

## 2. Model and Calculation method

The parallel-oriented magnetic trilayers for FePt/α-Fe adopted in this paper is shown in the inset of Fig. 1, where both soft and hard easy axes as well as the applied field $H$ are along the x axis and parallel to the film plane. The origin of the coordinate system is defined as the center of the hard/soft interface. For simplicity, all the films are assumed to spread to infinity, and the magnetostatic interaction could be ignored. As a result, the change of the magnetic moments with the applied field is within the film plane and the angles $\theta$ between the magnetic moments and the applied field in both phases depends only on the variable $z$. Due to the symmetry of the system, the calculations are performed only in the region defined by $-L^s/2 \leq z \leq L^h$, where the superscripts $h$ and $s$ denote the hard and soft phases, respectively.

According to Brown's micromagnetic theory[18-23], the total magnetic energy density per unit area for the trilayers can be expressed as

$$F = \int_0^{L^h} \left[ A^h \left( \frac{d\theta}{dz} \right)^2 + K^h \sin^2\theta - M_s^h H \cos\theta \right] dz$$
$$+ \int_{-\frac{L^s}{2}}^{0} \left[ A^s \left( \frac{d\theta}{dz} \right)^2 + K^s \sin^2\theta - M_s^s H \cos\theta \right] dz, \quad (1)$$

where $A$ denotes the exchange energy constant, $K$ is the anisotropy constant, $\theta$ is the angle between the magnetization and the applied field $H$, and $M_s$ is the spontaneous magnetization.

In this work the thickness of the hard layer is fixed as 10nm, which is much more than the Block wall width $\Delta^h$ ($\pi(A/K)^{1/2}$) for FePt (see table 1). As a result, the hard layer can be taken as infinitely thick in calculating the angle $\theta$[8].

The following boundary conditions[7, 8] can be obtained:

$$\theta\big|_{z=L^h} = 0, \frac{d\theta}{dz}\big|_{z=L^h} = 0;$$
$$\theta\big|_{z=0} = \theta^0; \quad (2)$$
$$\theta\big|_{z=-L^s/2} = \theta^s, \frac{d\theta}{dz}\big|_{z=-L^s/2} = 0,$$

$$A^s \frac{d\theta}{dz}\bigg|_{z=0^-} = A^h \frac{d\theta}{dz}\bigg|_{z=0^+}. \quad (3)$$

The two integral parts of the soft and hard phases in Eq. (1) are respectively substituted into the Euler-Lagrange equation $\frac{\partial F}{\partial \theta} = \frac{d}{dz}\frac{\partial F}{\partial \left(\frac{d\theta}{dz}\right)}$ within the variation method, and the following equations can be derived:

$$\frac{\pi(L^s/2+z)}{\Delta^s} = \int_\theta^{\theta^s} \frac{d\varphi}{\sqrt{(\sin^2\varphi - \sin^2\theta^s) - 2h^s(\cos\varphi - \cos\theta^s)}}, \quad (4)$$

$$\frac{\pi}{\Delta^h}\left(z\sqrt{1+h^h}\right) = -\ln\frac{\tan(\theta/2)\left[\sqrt{-1/h^h - 1} + \sqrt{-1/h^h - 1 - \tan^2(\theta^0/2)}\right]}{\tan(\theta^0/2)\left[\sqrt{-1/h^h - 1} + \sqrt{-1/h^h - 1 - \tan^2(\theta/2)}\right]}. \quad (5)$$

Here, $\theta^0$ and $\theta^s$ represent the angles between the magnetization and the applied field at the hard/soft interface ($z = 0$) and the center of the soft layer ($z = -L^s/2$), respectively, while $h^s = H/H_k^s$ and $h^h = H/H_k^h$ are the reduced applied fields for the soft and hard phases respectively, normalized by the corresponding anisotropy fields, $H_k^s = 2K^s/M_s^s$ and $H_k^h = 2K^h/M_s^h$.

Eqs. (4) and (5) are coupled by Eq. (3), which can be rewritten as:

$$\begin{aligned}A^s K^s\left[(\sin^2\theta^0 - \sin^2\theta^s) - 2h^s(\cos\theta^0 - \cos\theta^s)\right] \\ = A^h K^h\left[\sin^2\theta^0 - 2h^h(\cos\theta^0 - 1)\right].\end{aligned} \quad (6)$$

The subsequent calculation is based on Eqs. (4) - (6), with the material parameters extracted from Refs. [10, 12, 15] and listed in Table 1.

### 3. Results and Discussion

#### 3.1 Nucleation field

The nucleation field $H_N$ denotes the critical field[24] where the magnetization in the soft phase begins to deviate from the coherent state ($\theta = 0°$). At the nucleation point, the deviation from the coherent state is small (*i.e.*, $\theta \ll 1^0$)[10]. Thus the nucleation problem could be solved by the series expansion. Expanding Eqs. (4) and (6) and keeping only the two lowest terms, we have:

$$-h_N^s = 1 + \left[\frac{2\Delta^s}{\pi L^s}\cos^{-1}\left(\theta^0/\theta^s\right)\right]^2, \tag{7}$$

$$\left(\theta^s/\theta^0\right)^2 = 1 + \frac{A^h K^h \left(1 + h_N^h\right)}{A^s K^s \left(-h_N^s - 1\right)}. \tag{8}$$

Solving the above two linear equations, we obtain the nucleation field as an analytical function of the soft layer thickness $L^s$:

$$-h_N^s = 1 + \left[\frac{2\Delta^s}{\pi L^s}\tan^{-1}\frac{\sqrt{A^h K^h \left(1 + h_N^h\right)}}{\sqrt{A^s K^s \left(-h_N^s - 1\right)}}\right]^2, \tag{9}$$

where $h_N^s = -H_N/H_K^s$ is the reduced nucleation field.

Substituting the material parameters in table 1 into Eq. (9), we obtain the curves of the nucleation field as shown in Fig. 1. As $L^s$ increases from 0 to infinity, $H_N$ decreases smoothly from the anisotropy field of the hard phase to that of the soft phase. For thin soft layer, the nucleation field is strongly affected by the inherent properties of the hard layer, in particular, by $H_K^h$. When $L^s$ is larger than 40 nm, the curve is dominated by the parameters of the soft phase and the nucleation field approaches the anisotropy field of α-Fe.

## 3.2 Microscopic hysteresis loops for *θ$^s$* and *θ$^0$*

By solving Eqs. (4) and (6) numerically, we can obtain the hysteresis loops of $\theta^s$ and $\theta^0$, *i.e.*, the relationship of $\theta^s$ and $\theta^0$ with the applied field *H*.

Fig. 2 shows the microscopic hysteresis loops of the trilayers for various $L^s$. The curves of $\theta^s$ are shown in Fig. 2 (a), which can be divided to three steps: nucleation, motion and irreversible reversal of the domain wall, denoted by I, II and III, respectively. For $L^s$ = 20 nm, nucleation occurs at $H$ = -3.4 kOe, where $\theta^s$ jumps abruptly from 0° to $\theta_N^s$ (= 62°). This process is designated as section I, where the prototype of domain wall nucleates at the soft phase and the system changes from the coherent state to the incoherent one.

As $H$ decreases from -3.4 kOe to -4.4 kOe, $\theta^s$ rises gradually from $\theta_N^s$ to $\theta_P^s$ (= 136°), signifying an evolution and reversible motion of the domain wall from the soft to the hard phase, indicated by step II. This step quantifies the spring behavior in the hysteresis loops. Further decrease of $H$ will lead to another irreversible leap of $\theta^s$ from $\theta_P^s$ to 180°, corresponding to the pinning of the system and denoted as step III.

As $L^s$ increases, both $\theta_N^s$ and $\theta_P^s$ rise, signifying the extension of step I and the shrink of step III, so that the nucleation plays a more important role in the magnetic reversal process whilst the pinning becomes less important. Table 2 lists $\theta_N^s$ and $\theta_P^s$ shown in Fig. 2. One can see that step I ($\theta_N^s$) almost doubles whereas step III (180°-$\theta_P^s$) changes for the quarter as $L^s$ increases from 20 nm to 45 nm. In contrast, step II ($\theta_P^s - \theta_N^s$) does not have obvious change.

$\theta^0$ also experiences such three steps as shown in Fig. 2(b). However, step I for $\theta^0$ is much shorter whereas step III is much longer compared with the corresponding step for $\theta^s$. One can see from table 2, for the same $L^s$, $\theta_N^s$ is larger than $\theta_N^0$, indicating that $\theta^s$ responds to the applied field fast, which then drags $\theta^0$ through the exchange interaction. However, 180°-$\theta_P^s$ is much smaller than 180°-$\theta_P^0$, demonstrating that at section III, the main change of the magnetization is in the hard phase.

As $L^s$ increases, this discrepancy becomes more and more evident. In addition, the curves are pushed to the right, indicating the decline of both nucleation and pinning fields. For $L^s$ = 20 nm, $\theta_N^s$ is about 1.8 times of $\theta_N^0$ while 180°-$\theta_P^0$ is about 1.8 times of 180° -$\theta_P^s$, signifying that the two phases are still exchange-coupled quite well. However, as $L^s$ increases to 45 nm, $\theta_N^s$ is about three times of $\theta_N^0$ while 180°-$\theta_P^0$ is more than seven times of 180° -$\theta_P^s$, indicating that the two phases are decoupled.

**3.3 Macroscopic hysteresis loops**

The above microscopic hysteresis loops depict the underlying magnetic reversal mechanism well. However, they cannot directly illustrate the external magnetic properties of the material. To do this, we have obtained macroscopic hysteresis loops from Eqs. (4) - (6), shown in Fig. 3. As $L^s$ goes up, both nucleation and pinning fields decrease, consistent with the results in Fig. 2. As a result, the coercivity $H_c$ goes down with $L^s$ whereas the remanence rises.

The calculated $H_N$, $H_c$ and $H_P$ in Fig. 3 are highlighted in Fig. 4. For sufficiently small $L^s$, $H_N = H_P$ so that the three steps mentioned in section 3.2 are reduced to one single step and the hysteresis loop is rectangular. In this case, the coercivity also equals to the nucleation field and the coercivity mechanism is nucleation[10]. As $L^s$ increases, the curve of the pinning field detaches from that of the nucleation field and the single-step magnetic reversal expands to a standard three-step one, signifying the change of the coercivity mechanism from nucleation to pinning. The thickness where the coercivity mechanism alters is defined as the 1st critical thickness $L_{crit1}$ (6 nm). As $L^s$ goes up further, the coercivity is in between the nucleation and pinning fields and the coercivity mechanism changes from pinning to nucleation gradually. Similar to $L_{crit1}$, we define the thickness where the coercivity detaches from the pinning field as the 2nd critical thickness and denoted as $L_{crit2}$. According to Fig. 4, $L_{crit2}$ is 20 nm, which is much larger than $L_{crit1}$.

These changes of coercivity mechanism are somewhat different from those obtained in Refs. [10, 11], where the coercivity mechanism changes unidirectionally from nucleation to pinning as $L^s$ increases. Close analyses show that in Refs. [10, 11], the hard layer thickness $L^h$ is set as infinite. The main contribution to the magnetization of the material is from that of the hard layer so that the coercivity cannot be smaller than the pinning field, where the magnetization in the hard layer reversals. This situation holds for $L^s < L_{crit2}$, when $L^h$ is adopted as 10 nm. However, for larger $L^s$, the hysteresis loops are dominated by the magnetic behaviors of the soft layer and the coercivity mechanism changes to the nucleation.

The first critical thickness can be derived analytically by considering the fourth order term of $\theta$ in Eq. (1)[10]. The energy change at nucleation is given by:

$$\Delta E = B\left(v \tan v - \sin^2 v \cos 2v - \frac{2}{\eta}\cos^4 v\right)\theta^4, \qquad (10)$$

where $B$ is positive, $v = \frac{\pi L^s}{2\Delta^s}\sqrt{-h^s}$ (11) and $\eta = \frac{A^s M^s}{A^h M^h}$ (12).

Setting Eq. (10) = 0, we can easily determine the first critical thickness $L_{crit1}$ for various materials, shown in Table 3:

As shown in Table 3, our calculated first critical thicknesses $L_{crit1}$ are somewhat smaller than the corresponding Bloch wall width of the hard phase given in table 1, consistent with available numerical results[10, 12].

From the above discussions, one can see that many magnetic properties, such as the critical thickness and the nucleation fields, rely largely on the crystalline anisotropy of the hard phase. To have systematic information on the influence of the $K^h$ on $L_{crit1}$, the first critical thickness has been calculated as a function of $K^h$, as shown in Fig. 5. The calculated $L_{crit1}$ is inversely proportional to the square root of $K^h$.

To understand this we have done the following derivations. Substituting Eq. (9) into Eq. (11), we obtain the relationship between the nucleation field at the first critical thickness and the crystalline anisotropy of the hard phase:

$$\frac{K^h}{H_N} = \frac{M_s^h}{2}\left(1+\eta \tan^2 v\right), \quad (13)$$

where the approximation of $(-h_N^s - 1) \approx -h_N^s$ is adopted because $-h_N^s \gg 1$.

Substituting Eq. (13) into Eq. (9), the explicit formula for the first critical thickness $L_{crit1}$ turns out:

$$L_{crit1} = \sqrt{\frac{A^s M_s^h}{M_s^s}\left(1+\eta \tan^2 v\right)} \cdot \frac{2v}{\sqrt{K^h}}. \quad (14)$$

Eq. (14) clearly demonstrates that the first critical thickness scales with $(K^h)^{-1/2}$, as shown in Fig. 5. In addition, the slope of the line is determined by $A^h$, $A^s$, $M_s^h$ and $M_s^s$ according to Eq. (14).

## 4. Conclusions

The magnetic reversal processes in parallel-oriented hard/soft trilayers have been discussed within a self-contained micromagnetic model.

The nucleation field decreases monotonically with the increase of $L^s$. For thin soft layer, the nucleation field is dominated by $H_K^h$. For $L^s > 40$ nm, the curve is subject to the parameters of the soft phase and the corresponding nucleation field approaches the anisotropy field of α-Fe.

The magnetizations in the soft phase respond to the applied field fast, which then drag those in the interface through the exchange interaction. As $L^s$ goes up, both nucleation and pinning fields falls. In the meantime, the coercivity reduces whereas the remanence rises. The coercivity mechanism changes from nucleation to pinning, and finally to nucleation again as $L^s$ increases. The first critical thickness $L_{crit1}$ at which the coercivity mechanism changes from nucleation to pinning can be determined analytically, which is less than the corresponding Bloch wall width of the hard phase. Similar to the Bloch wall width, the calculated $L_{crit1}$ is inversely proportional to the square root of the crystalline anisotropy of the hard phase, consistent with available numerical results.

## Acknowledgement

This work was supported by the National Natural Science Foundation of China (Grant No. 10747007) and the Scientific Research Foundation for Returned Overseas Chinese Scholars, State Education Ministry.

**References**

1. E.F. Kneller, R. Hawig, IEEE Trans. Magn., **27,** 3588-3560 (1991).

2. R. Skomski and J. M. D. Coey, Phys. Rev. B **48**, 15812 (1993).

3. S. S. Yan, W. J. Liu, J. L. Weston, G. Zangari, and J. A. Barnard, Phys. Rev. B **63**, 174415 (2001).

4. S. S. Yan, M. Elkawni, D. S. Li, H. Garmestani, J. P. Liu, J. L. Weston and G. Zangari, J. Appl. Phys. **94,** 4535 (2003).

5. S. G. Wang, R. C. C. Ward, G. X. Du, X. F. Han, C. Wang, and A. Kohn, Phys. Rev. B **78**, 180411(R) (2008).

6. S. G. Wang, C. Wang, A. Kohn, S. Lee, J. P. Goff, L. J. Singh, Z. H. Barber and R. C. C. Ward, J. Appl. Phys.**101**, 09D103 (2007).

7. G. P. Zhao, N. Bo, H. W. Zhang, Y. P. Feng, and Y. Deng, J. Appl. Phys. **107**, 083907 (2010).

8. C. W. Xian, G. P. Zhao, Q. X. Zhang, and J. S. Xu, Acta. Phys. Sin. **58**, 3509-3514 (2009, in Chinese).

9. T. Leineweber and H. Kronmüller, J. Magn. Magn. Mater. **176**, 145 (1997).

10. G. P. Zhao and X. L. Wang, Phys. Rev. B **74**, 012409 (2006)

11. G. P. Zhao, M.G. Zhao, H. S. Lim, Y. P. Feng, C. K. Ong, Appl. Phys. Lett. **87**, 162513 (2005).

12. G. Asti, M. Solzi, M. Ghidini, and F. M. Neri, Phys. Rev. B **69**, 174401 (2004).

13. G. Asti, M. Ghidini, R. Pellicelli, C. Pernechele, M. Solzi, F. Albertini, F. Casoli, S. Fabbrici, and L. Pareti, Phys. Rev. B **73**, 094406 (2006).

14. M. G. Pini, and A. Rettori, Phys. Rev. B **60**, 3414 (1999).

15. K. Yu. Guslienko, O. Chubykalo-Fesenko, O. Mryasov, R. Chantrell, and D. Weller, Phys. Rev. B **70**, 104405 (2004).

16. Y. K. Takahashi, T. O. Seki, K. Hono, T. Shima and K. Takanashi, J. Appl. Phys. **96,** 475-481 (2004).

17. G. P. Zhao, L. Chen, C. W. Huang, and Y. P. Feng, J. Magn. Magn. Mater. **321**, 2322 (2009).

18. W. F. Brown, Jr., J. Appl. Phys. **33**: 3026-3031 (1962).

19. W. F. Brown, Jr., Micromagnetics (Wiley Interscience, New York, 1963), p. 72.

20. W. F. Brown, Jr., Am. J. Phys. **28**, 542-551 (1960).

21. C. E. Johnson, Jr. and W. F. Brown, Jr., J. Appl. Phys. **32**, 243S-344S (1961).

22. A. Aharoni, S. Shtrikman, Phys. Rev. **109,** 1522 (1958).

23. G. P. Zhao, L. Chen, C. W. Huang , N. L. Guo, Y. P. Feng, Solid State Communications **150**, 1486-1488 (2010).

24. W. F. Brown, Jr., Rev. Mod. Phys. **17**, 15 (1945).

**Figure captions**

Fig.1: Calculated nucleation field for a FePt/α-Fe/FePt trilayer system (shown in the inset) according to Eq. (9).

Fig. 2: Calculated microscopic hysteresis loops of the trilayers for $\theta^s$ and $\theta^0$.

Fig. 3: Calculated macroscopic hysteresis loops of the trilayers for various $L^s$.

Fig. 4: Calculated $L^s$ dependent nucleation, pinning and coercive fields.

Fig. 5: Calculated first critical thickness $L_{crit1}$ as a function of the crystalline anisotropy of the hard phase in a FePt/α-Fe trilayer system.

Table 1: Magnetic parameters for the hard and soft phases.

| Materials | $K(\times 10^7 \text{erg/cm}^3)$ | $M(\times 10^3 \text{emu/cm}^3)$ | $A(\times 10^{-7} \text{erg/cm})$ | $H_K$ (kOe) | $\Delta$ (nm) |
|---|---|---|---|---|---|
| FePt | 2 | 1.10 | 8 | 36.4 | 6.28 |
| α-Fe | 0.046 | 1.71 | 25 | 0.54 | 73.16 |
| Nd$_2$Fe$_{14}$B | 4.3 | 1.28 | 7.7 | 67.2 | 4.2 |
| SmCo$_5$ | 17.1 | 0.84 | 12 | 407 | 2.6 |
| Sm$_2$Fe$_{17}$N$_3$ | 12 | 1.23 | 10.7 | 195 | 3.0 |
| Co | 0.43 | 1.43 | 10.3 | 6.0 | 15.4 |
| Fe | 0.0001 | 1.7 | 28 | 0.001177 | 1662.4 |
| Sm-Co | 5 | 0.55 | 12 | 181.8 | 4.87 |

Table 2: Calculated heights of $\theta^s$ and $\theta^0$ at steps I, II and III where subscripts $N$ and $P$ denote the nucleation and pinning, respectively.

| $L^s$ (nm) | $\theta^s_P$ | $\theta^s_N$ | $\theta^s_P - \theta^s_N$ | $\theta^0_P$ | $\theta^0_N$ | $\theta^0_P - \theta^0_N$ |
|---|---|---|---|---|---|---|
| 20 | 135.6° | 62.0° | 73.6° | 98.2° | 34.6° | 63.6° |
| 25 | 149.6° | 71.5° | 78.1° | 109.9° | 35.7° | 74.2° |
| 30 | 156.0° | 83.3° | 72.7° | 109.0° | 37.6° | 71.4° |
| 35 | 161.9° | 97.1° | 64.8° | 113.9° | 40.3° | 73.6° |
| 40 | 167.4° | 108.8° | 58.6° | 112.9° | 41.9° | 71.0° |
| 45 | 170.6° | 119.2° | 51.4° | 112.6° | 42.8° | 69.8° |

Table 3: Calculated first critical thickness $L_{crit1}$ and corresponding nucleation field for various magnetic materials.

| Materials | $L_{crit1}$(nm) | Nucleation field(kOe) |
|---|---|---|
| FePt /α-Fe | 5.13 | 13.55 |
| $Nd_2Fe_{14}B$/α-Fe | 3.84 | 25.29 |
| $SmCo_5$/Co | 1.13 | 211.89 |
| $Sm_2Fe_{17}N_3$/α-Fe | 2.30 | 78.87 |
| Sm-Co/ Fe | 2.48 | 55.79 |

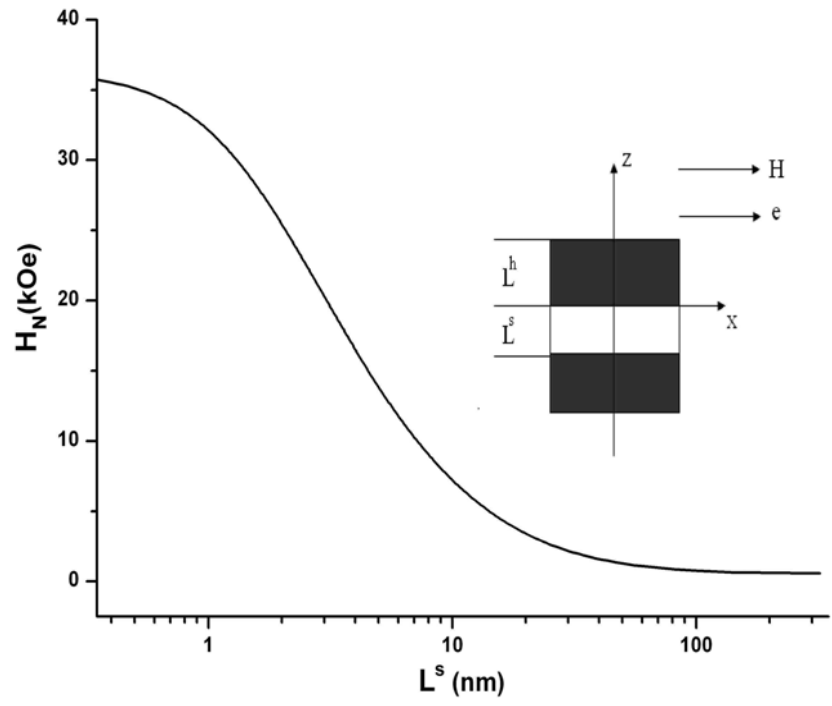

**Fig.1**

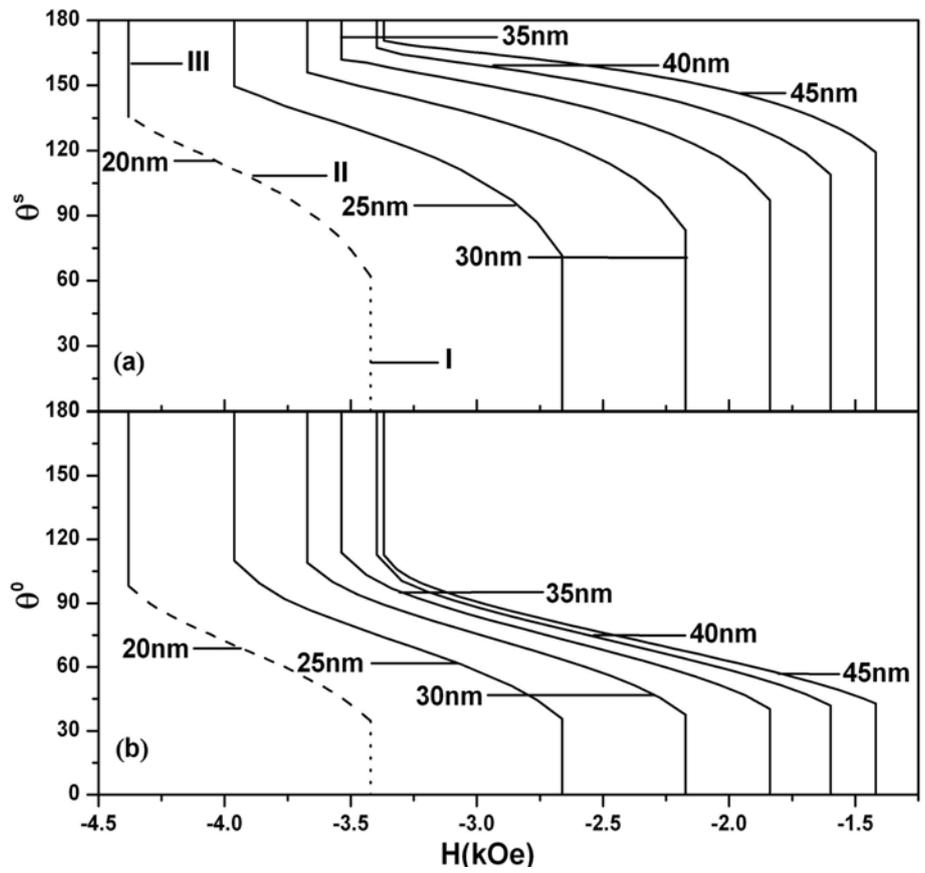

**Fig.2**

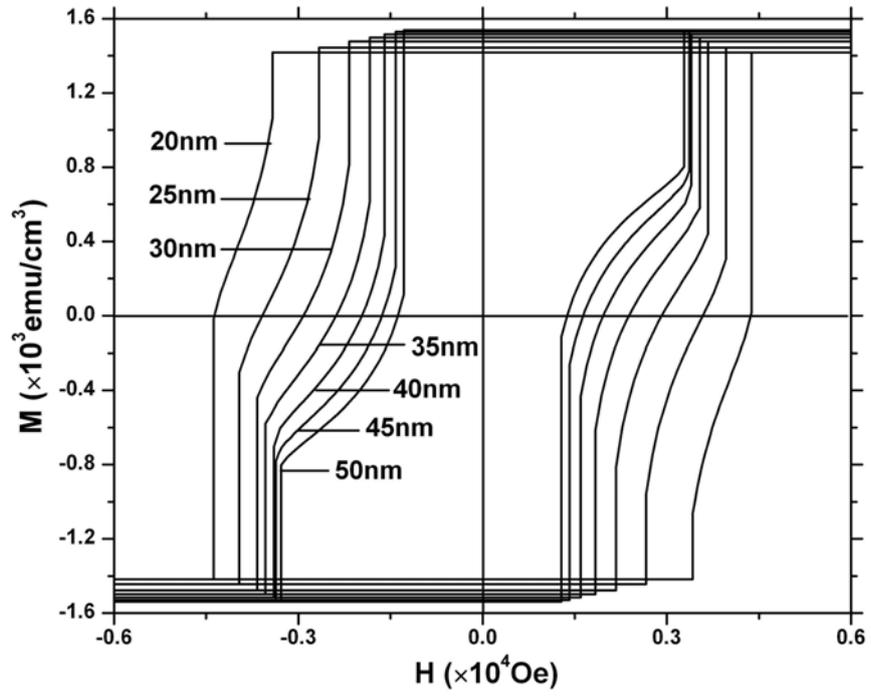

**Fig.3**

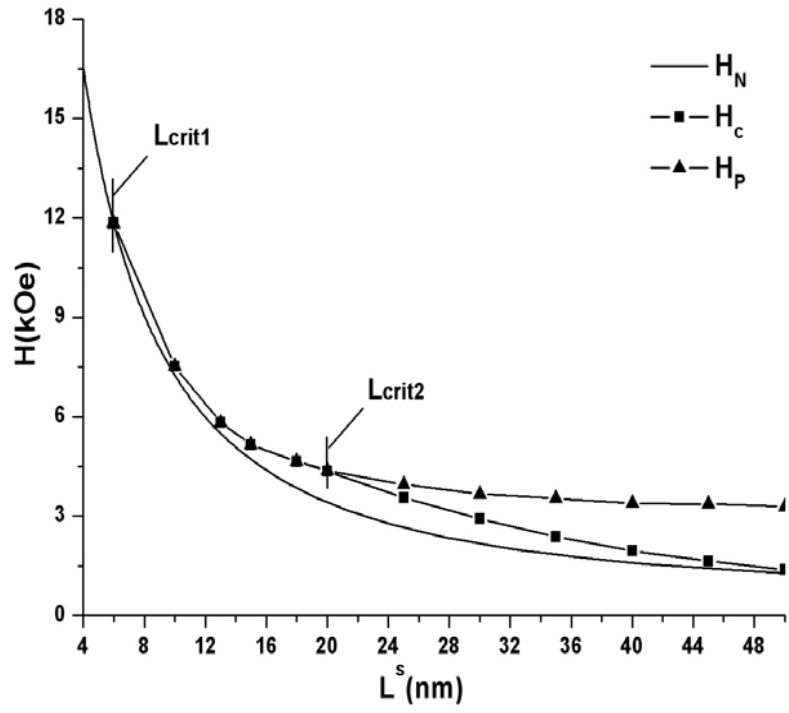

**Fig.4**

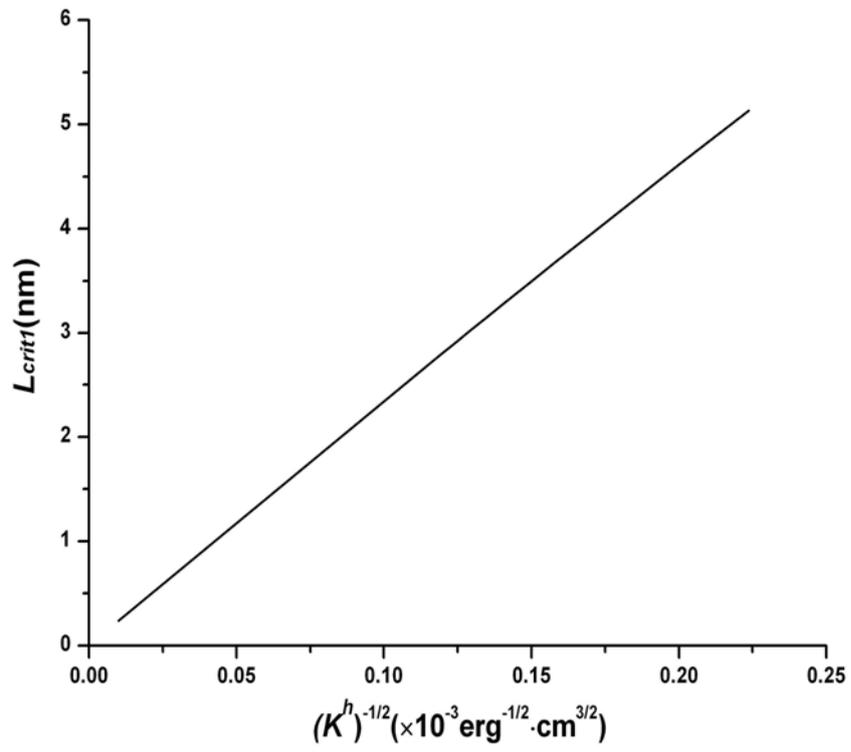

**Fig.5**